# Polarization-controlled modulation doping of a ferroelectric from first principles

Xiaohui Liu[1*] Evgeny Y. Tsymbal[2] and Karin M. Rabe[1]

[1]*Department of Physics and Astronomy, Rutgers University, Piscataway, 08854, USA*

[2] *Department of Physics and Astronomy & Nebraska Center for Materials and Nanoscience, University of Nebraska, Lincoln, Nebraska 68588-0299, USA*

In a ferroelectric field effect transistor (FeFET), it is generally assumed that the ferroelectric gate plays a purely electrostatic role. Recently it has been shown that in some cases, which could be called "active FeFETs", electronic states in the ferroelectric contribute to the device conductance as the result of a modulation doping effect in which carriers are transferred from the channel into the ferroelectric layers near the interface. Here we report first-principles calculations and model analysis to elucidate the various aspects of this mechanism and to provide guidance in materials choices and interface termination for optimizing the on-off ratio, using $BaTiO_3$/n-$SrTiO_3$ and $PbTiO_3$/n-$SrTiO_3$ as prototypical systems. It is shown that the modulation doping is substantial in both cases, and that extension of an electrostatic model developed in previous work provides a good description of the transferred charge distribution. This model can be used to suggest additional materials heterostructures for the design of active FeFETs.

In a field-effect transistor, the conductance of the channel is modulated by a voltage applied between the gate and the base. A ferroelectric field-effect transistor (FeFET) is switched between high-conductance ON and low-conductance OFF states by switching the spontaneous polarization of the ferroelectric gate [1, 2]. If the role of the ferroelectric gate is purely electrostatic, then the difference in conductance between the up and down polarization states results from the change in channel carrier density that screens the depolarization field in the ferroelectric, and the concomitant change in the density of states at the Fermi level (Figure 1(a) and (b)). This change in carrier density is largest within a screening length of the interface. The fractional change in conductance, the "on-off ratio," is greatest when the carrier density of the bulk material of the channel is low, as in a doped semiconductor, complex oxide, or graphene sheet [3]. For example, modulation of the conductance by 300% was found in a $PbZr_{0.2}Ti_{0.8}O_3$/$La_{0.7}Ca_{0.3}MnO_3$ heterostructure [4] and by more than 600% in $PbZr_{0.2}Ti_{0.8}O_3$/graphene FeFETs [5].

Recent first-principles studies of ferroelectric heterostructures suggest that in some cases the modulation of the conductance is not solely due to the change in carrier density in the channel material, but can include active involvement of the ferroelectric, with significant contributions from interfacial electronic reconstruction [6, 7, 8, 9, 10] opening new high-conductivity channels in one polarization state (Figure 1(c)). The analysis of observed changes of conductance driven by ferroelectric polarization switching at a ferroelectric - complex oxide interface $PbZr_{0.2}Ti_{0.8}O_3$/$LaNiO_3$ [11] showed that a new conducting channel opened in the interface PbO layer for polarization pointing into the interface. First-principles calculations for a $SrRuO_3$/$BaTiO_3$/n-$SrTiO_3$ ferroelectric tunnel junction showed metallization of two layers of $BaTiO_3$ at the $BaTiO_3$/n-$SrTiO_3$ interface [12], which suggested a mechanism for large tunneling electroresistance in which both the barrier height and effective barrier width change as the polarization is switched. For the same heterostructure considered as a FeFET, this metallization indicates the opening of a new conducting channel in the ferroelectric interface layers.

This behavior offers a promising avenue to enhance the on-off ratio in a FeFET by focusing on active involvement of the ferroelectric gate. The transfer of charge carriers into the ferroelectric gate via modulation doping is determined by the choice of materials and the terminations at the interface. The contribution to the conductance from the transferred carriers can be made larger than that of the carriers in the doped semiconductor by choice of a ferroelectric material with a high mobility for added carriers and the reduction of scattering by impurity dopants, which reside in the doped semiconductor.

In this paper, we report first-principles calculations and model analysis to elucidate the various aspects of this mechanism for conductivity switching and to provide guidance in materials choices and interface termination for optimizing the on-off ratio. We use $BaTiO_3$/n-$SrTiO_3$ and $PbTiO_3$/n-$SrTiO_3$ as prototypical systems. We show that the modulation doping is substantial in both cases and apply an extension of an electrostatic model developed in previous work [12] to obtain a good description of the transferred charge distribution. This model can be used to suggest additional materials heterostructures for the design of active FeFETs.

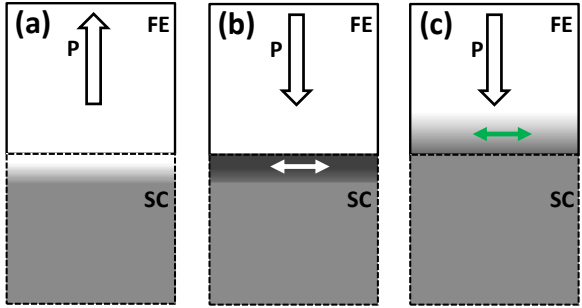

Figure 1. A schematic of the effect of polarization direction on the conductance of a ferroelectric/doped-semiconductor heterostructure. (a) For one choice of polarization direction, the majority charge carriers in the doped semiconductor are pushed away from the interface, reducing the conductivity and switching the device to the off state. (b) When the polarization direction is reversed, if the role of the ferroelectric is purely electrostatic, increase in carrier density and concomitant increase in the density of states at the Fermi level occurs only in the channel material within a screening length of the interface. (c) For an active ferroelectric gate, the carrier density also becomes nonzero in the ferroelectric layers adjacent to the interface through modulation doping, opening a new conducting channel (indicated by the green double-headed arrow) in the ferroelectric interface layers.

First-principles calculations were performed using Quantum ESPRESSO [13] within the local density approximation (LDA) and LDA+U. Ultrasoft pseudopotentials with plane-wave basis limited by a cutoff energy of 40 Ry are used, including 10 valence electrons for $Sr(4s^24p^65s^2)$, 10 for $Ba(5s^25p^66s^2)$, 11 for $Ti(3s^23p^64s^23d^1)$, 6 for $O(2s^22p^4)$. Nonzero U was included using the linear response method [14]. The Brillouin zone is sampled by a 6×6×1 mesh of k points. Additional details are given in the Supplemental Material.

We considered 1x1 $(SrRuO_3)_5/(ATiO_3)_8/(n-SrTiO_3)_{16}/(ATiO_3)_8/(SrRuO_3)_5$ (A = Ba, Pb) supercells stacked along the [001] direction with mirror symmetry around the central SrO atomic plane. This supercell geometry avoids direct contact between the two electrode materials, $SrRuO_3$ and n-$SrTiO_3$, and ensures full compatibility of arbitrary polarization of the $BaTiO_3$ layers with periodic boundary conditions. As the role of $SrRuO_3$ in this system is only as a top electrode and carrier reservoir, we treat it as a nonmagnetic material with no rotational distortions. At the $SrRuO_3/BaTiO_3$ interfaces, $BaTiO_3$ is terminated with $TiO_2$. The in-plane lattice constant of the supercell is constrained to the calculated LDA lattice constant of $SrTiO_3$, $a = 3.851$ Å, which corresponds to an in-plane strain of about -2.1% on $BaTiO_3$ and -0.14% on $PbTiO_3$. This epitaxial constraint stabilizes $BaTiO_3$ in the P4mm tetragonal phase with a spontaneous polarization of 40.9 μC/cm$^2$ and $c$ parameter of 4.101 Å. The tetragonal $PbTiO_3$ has a spontaneous polarization 80.9 μC/cm$^2$ and $c$ parameter 4.032 Å. An electron concentration of 0.09 per formula unit is produced via a scaling of the oxygen pseudopotential in the $SrTiO_3$ layers (6.03 valence electrons for $O(2s^22p^{4.03})$.). The $BaTiO_3$/n-$SrTiO_3$ interfaces are terminated with doped $TiO_2$. The atomic positions are relaxed until forces are converged to less than 20 meV/Å on each atom, with the supercell constrained to be tetragonal so that only the c parameter is allowed to relax. Following Ref. [12], the layer-by-layer density of states is obtained by recomputing the electronic states of the relaxed structure with U = 5 eV for the Ti d-states in the $BaTiO_3$ layer to correct artifacts arising from the LDA underestimate of the band gap.

The system is found to have two locally stable states: one in which the polarization of the $BaTiO_3$ layer points away from the $BaTiO_3$/n-$SrTiO_3$ interface, and one in which the polarization points into the interface. As previously discussed [12], in the former case, the depolarization field is screened by a combination of depletion of electrons and polar lattice distortions in the region of n-$SrTiO_3$ near the interface. The $BaTiO_3$ layers are insulating, and in addition, the conduction band minimum in the $SrTiO_3$ layers adjacent to the interface is pushed up above the Fermi level, so that these layers are insulating as well. For polarization pointing into the interface, in addition to the accumulation of electrons and polar lattice distortions in the interface region of n-$SrTiO_3$, electrons are transferred into the interface layers of $BaTiO_3$, making a substantial additional contribution to the screening. The downward bending of the bands metallizes the ferroelectric interface layers. In addition, the free carriers reduce their polar distortion, consistent with experimental and theoretical results that show that the polar distortion of bulk $BaTiO_3$ is reduced by electron doping through oxygen vacancy or substitution of Ba by La but remains nonzero up to a La concentration of 0.15 [15,16,17]. Finally, we note that in this geometry, the two polarization states are inequivalent, with the magnitude of the polarization pointing into the interface being smaller than that pointing away from the interface, due to the dissimilar electrodes (n-$SrTiO_3$ and $SrRuO_3$).

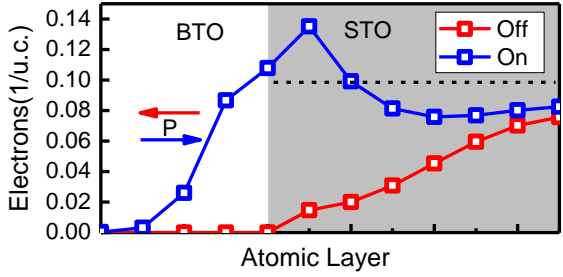

FIG. 2. Excess electrons in each unit cell layer for two polarization directions, obtained by integrating the occupation of the local density of states above the conduction band minimum presented in Fig. 2 of Ref [12]. Arrows indicate the direction of polarization.

The excess electron density profile is computed by integrating the occupation of the local density of states above the conduction band minimum in each unit cell layer. The profile from the middle layer of BaTiO$_3$ to the midpoint of the n-SrTiO$_3$ layer is shown in Fig. 2. When polarization is pointing away from the interface, the excess electron density in n-SrTiO$_3$ is reduced below the doping level of 0.09 electrons/u.c even well away from the interface, producing a wide depletion region. In the supercell considered, these electrons are transferred to the SrRuO$_3$ layer. When polarization is pointing into the interface, conduction band levels are occupied in the two layers of BaTiO$_3$ at the interface, and the excess election density increases above the doping level at the interface and in the two adjacent layers of n-SrTiO$_3$. In addition to transfer of electrons from the SrRuO$_3$ electrode (not shown), we note that electrons are also transferred from the n-SrTiO$_3$ layers away from the interface.

In Figure 3, we present plots of the spatial dependence of the density of states near the Fermi level, analogous to those presented in Ref. [11]. When the polarization points away from the BaTiO$_3$/n-SrTiO$_3$ interface, the density of states near the Fermi level in the first three layers of n-SrTiO$_3$ is dramatically reduced. When polarization points into the BaTiO$_3$/n-SrTiO$_3$ interface, the density of states near the Fermi level in the interface layers of n-SrTiO$_3$ increases slightly and the two interface layers of BaTiO$_3$ are metallized.

In Figure 4, we show the effect of the polarization direction on the conduction band states in BaTiO$_3$ near the interface by projecting the Ti d bands of BaTiO$_3$ in the first six unit cell layers [18]. When the polarization points away from the BaTiO$_3$/n-SrTiO$_3$ interface, these states are above the Fermi level, as shown by Figure 4(a). When polarization points into the BaTiO$_3$/n-SrTiO$_3$ interface, these states are shifted down in energy, and two bands cross the Fermi level as shown in Figure 4 (b), resulting in metallic character of the layers.

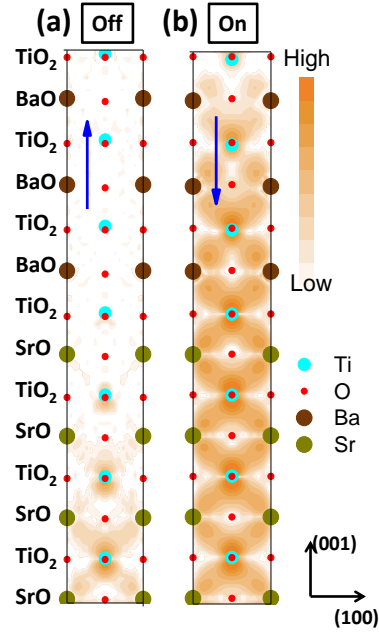

FIG. 3. A 2D projection of the spatial dependence of the local density of electronic states derived from first-principles calculations integrated within $\pm k_B T$ eV of the Fermi level with T=300K near the interface BaTiO$_3$/n-SrTiO$_3$ for (a) polarization pointing away from the interface, and (b) polarization pointing into the interface.

To understand how the direction of the ferroelectric polarization affects the carrier distribution near the interface, we extend the model used in our previous paper to analyze the SrRuO$_3$/BaTiO$_3$/n-SrTiO$_3$ system. This electrostatic model describes each electrode (SrRuO$_3$ and n-SrTiO$_3$) by its screening length, relative dielectric constant, and Fermi level relative to the vacuum reference, and the ferroelectric by its polarization, conduction band minimum relative to the vacuum reference, and density of states near the conduction band minimum. Self-consistent solution of the model, described in the Supplemental Material, yields an electrostatic potential profile, which also specifies the bending of the bands, and a charge density profile. The electrostatic potential at each ferroelectric/electrode interface, called the screening potential $\varphi$, increases with the screening length of the electrode and the spontaneous polarization of the ferroelectric. The carrier distribution is determined by the relative values of $\varphi$ and $\Delta\Phi$, the difference between the conduction band minimum of the ferroelectric and the Fermi level of the doped semiconductor. If $\Delta\Phi > 0$ and $\varphi$ is smaller than $\Delta\Phi$, then no carriers are transferred into the ferroelectric in either polarization state. If the barrier $\Delta\Phi$ is smaller than $\varphi$, then for polarization pointing into the interface, the electrostatic potential lowers the conduction band of the ferroelectric near the interface below the Fermi level, and electrons are transferred into the interface layers of the ferroelectric and contribute to the conductance. This transfer is promoted by a small difference between the conduction band minimum of the ferroelectric and the Fermi level of the doped semiconductor, and by a large spontaneous polarization.

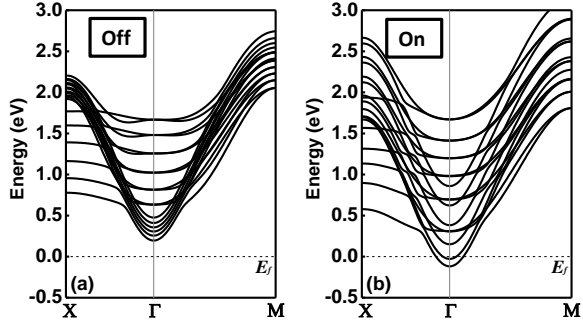

FIG. 4. Bands structure of the (a) Off state and (b) On state projected on $BaTiO_3$ layers near the interface. Dash line indicates the Fermi level of the heterostructure.

The transfer of electrons in the $SrRuO_3/BaTiO_3/n$-$SrTiO_3$ system can be readily understood within this model. An estimate for $\Delta\Phi$ can be obtained from the measured electron affinities of $BaTiO_3$ and $SrTiO_3$. These are both about 4.0 eV, giving a value of $\Delta\Phi$ close to zero. A first-principles estimate for the band alignment can be obtained by lining up the centers of the oxygen 2p bands, as described in Refs. [19,20]. With LDA, the conduction bands minima are 2.90 eV and 3.10 eV with respect to the lined up oxygen 2p center for $SrTiO_3$ and $BaTiO_3$, giving a value of $\Delta\Phi$ of 0.2 eV. We performed calculations with the Heyd-Scuseria-Ernzerhof (HSE) hybrid functional.[21] These calculations give band gaps which match experiment very well. By lining up the oxygen 2p center, we find that the conduction bands minima are 4.87 eV and 4.90 eV for $SrTiO_3$ and $BaTiO_3$, giving a value of $\Delta\Phi$ of 0.03 eV. In the previous study, the model showed a shift of bands about 1.7 eV between the two polarization states.[3]

$PbTiO_3$ has a smaller electron affinity than $BaTiO_3$ (about 3.5 eV), which would increase $\Delta\Phi$, decreasing this effect. An HSE calculation analogous to that above gives the conduction band minimum of $PbTiO_3$ as 5.30 eV and $\Delta\Phi$ of 0.43 eV (the LDA calculation gives conduction band minimum 3.20 eV and $\Delta\Phi$ of 0.30 eV). However, $PbTiO_3$ also has a larger polarization than $BaTiO_3$, which increases the screening potential at the $PbTiO_3/n$-$SrTiO_3$ interface as shown in the supplementary materials, and would increase the effect. By performing first-principles calculation on a $SrRuO_3/PbTiO_3/n$-$SrTiO_3$ heterostructure, described in detail in the Supplemental Material, we find that the net effect is comparable to what was found in $SrRuO_3/PbTiO_3/n$-$SrTiO_3$. Specifically, our calculation indicates the metallization of more than two layers of $PbTiO_3$ near the $PbTiO_3/n$-$SrTiO_3$ interface when polarization is pointing into the interface.

The central role of $\Delta\Phi$ also suggests increasing or decreasing the degree of metallization by modifications of the interface that change the band alignment. Previous theoretical and experimental studies have explored various types of interface engineering. For example, it was demonstrated that stoichiometry of the interfacial $La_{1-x}Sr_xO$ layer at the $La_{0.7}Sr_{0.3}MnO_3/SrTiO_3$ interface can be used to control the Schottky barrier height [22, 23]. It was also shown that A-site composition allows tuning of the band offset at the $Ba_{1-x}Sr_xTiO_3/Ge$ interface [24]

The conductivity of (Ba, La)$TiO_3$ has been measured at room temperature with different doping levels, yielding a value of mobility of several $cm^2V^{-1}s^{-1}$. At room temperature, the scattering is dominated by phonons. At low temperatures, the separation of the free carriers in the ferroelectric interface layer from the impurity atoms in the doped semiconductor should result in substantially enhanced on-state conductivity and on/off ratio. As is pointed out in [25], tuning the strain could further enhance the mobility of the system. In fact, the epitaxial growth of $BaTiO_3$ and $PbTiO_3$ on $SrTiO_3$ introduce considerable strain on ferroelectric. Therefore, we can still expect the additional channel in ferroelectric has higher conductivity.

In summary, we have investigated the active involvement of the ferroelectric gate in the conductance of a FeFET from first principles calculations and modeling. We showed that this involvement, based on polarization-dependent modulation doping, is promoted by minimizing the work function difference between the ferroelectric and the doped semiconductor and maximizing the ferroelectric polarization. Enhancement of the on-off ratio could thus be achieved with use of a high-mobility ferroelectric. Our first-principles results for $BaTiO_3/n$-$SrTiO_3$ and $PbTiO_3/n$-$SrTiO_3$ illustrate the mechanism and are practical starting points for experimental investigation of this effect.

We thank C. H. Ahn, S. Ismail-Beigi, D. R. Hamann, D. Vanderbilt and Cyrus Dreyer for valuable discussion. First-principles calculations were performed on the Rutgers University Parallel Computer (RUPC) and the Nebraska Holland Computing Center cluster. This work

was supported by ONR N00014-14-1-0613 and NSF DMR-1334428.

* phyliuxiaohui@gmail.com

# Supplementary Material: **Polarization-controlled modulation doping of a ferroelectric from first principles**

Xiaohui Liu[1] Evgeny Y. Tsymbal[2] and Karin M. Rabe[1]

[1] Department of Physics and Astronomy, Rutgers University, Piscataway, 08854, USA
[2] Department of Physics and Astronomy & Nebraska Center for Materials and Nanoscience, University of Nebraska, Lincoln, Nebraska 68588-0299, USA

**A: Band bending in the ferroelectric near the interface**

Consider a ferroelectric capacitor with a nonzero spontaneous polarization. Due to the imperfect screening of the electrodes, this polarization results in nonzero screening potentials near the interfaces. Because the potential must be continuous, the imperfect screening leads to the bending of the bands of the ferroelectric up or down, depending on the direction of the ferroelectric polarization. We compare the band bending to the barrier height, defined as the difference between the work function of the electrode and the conduction band minimum of the ferroelectric. For good metal electrodes, the screening length is less than 0.1 nanometers and the potential drop near the interface is smaller than the barrier height. However, if the screening length is large enough, as might happen when the electrode is a doped semiconductor, the screening potential near the interface could be larger than the barrier height, leading to charge transfer between the electrode and the ferroelectric.

We model this effect using the approximation of the Thomas-Fermi model of screening, In that case, the screening potentials of the two (left and right) interfaces are given by:[1]

$$\varphi_l(z) = \frac{\sigma_s \delta_l e^{-|z|/\delta_l}}{\varepsilon_0} \quad \text{and} \quad \varphi_r(z) = \frac{\sigma_s \delta_r e^{-|z-d|/\delta_r}}{\varepsilon_0}$$

where $\sigma_s$ is the magnitude of the screening charge density, given by $\sigma_s = \frac{dP}{\varepsilon(\delta_l+\delta_r)+d}$.

We fix the screening length of the left interface to be $\delta_l = 0.1\ nm$ and find the screening potential of the right interface as a function of screening length $\delta_r$. We take the thickness of the ferroelectric layer to be 5 nm and the relative dielectric constant of the ferroelectric to be 100, which are typical values for ferroelectric thin films [2].

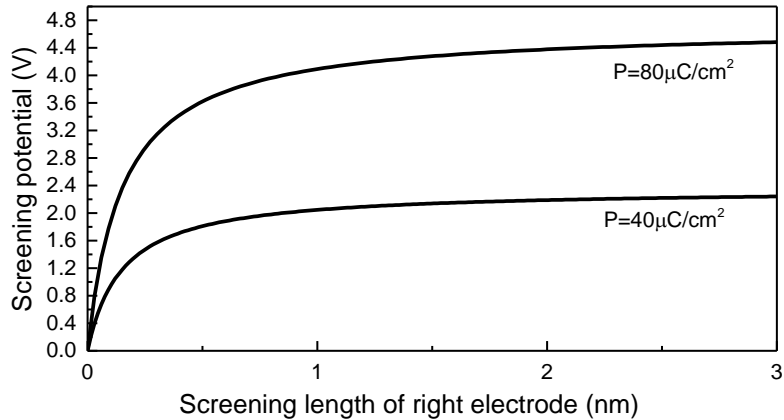

Fig. 1s Screening potential as a function of screening length of the right electrode, with screening length of the left electrode fixed at 0.1 nm.

Fig. 1s shows the calculated dependence of the screening potential on screening length in two cases, one with polarization 40 μC/cm$^2$ and the other with polarization 80 μC/cm$^2$, corresponding to compressively strained BaTiO$_3$ and PbTiO$_3$, respectively. We see that the screening length has a dramatic effect on the screening potential. If the screening length of the right electrode is larger than 1 nm, the screening potential is more than 1 eV. Such a large screening potential has a substantial effect on the electronic structure at the interface. When this screening potential is larger than the barrier height at the interface, then the conduction bands of the ferroelectric layer bend down below the Fermi level of the system for polarization pointing into the electrode. This could happen if the electrode is a doped semiconductor, since the screening length would be much larger than that of a good metal.

By calculating the electrostatic potential profiles of the SrRuO$_3$/BaTiO$_3$/n-SrTiO$_3$ system for each polarization direction, we can estimate how much the conduction bands of BaTiO$_3$ bend down when the polarization is pointing into the n-SrTiO$_3$ electrode. As shown in Fig. 2s, the electrostatic potential energy at the interface of BaTiO$_3$/n-SrTiO$_3$ is shifted up and down by about 0.64 eV. We divide by two to get 0.32 eV as the screening potential induced by polarization. Considering that the barrier height between BaTiO$_3$ and n-SrTiO$_3$ is about 0.1 eV, as discussed in the main text, this screening potential is large enough to bend the conduction band of BaTiO$_3$ below the Fermi level and metallize the BaTiO$_3$ layers near the interface.

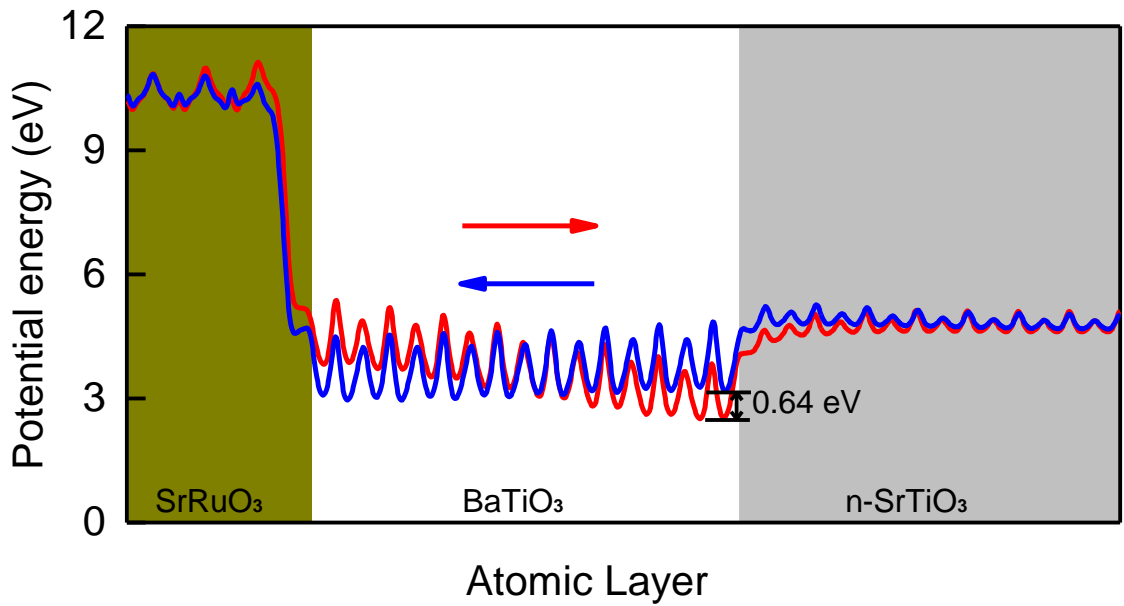

Fig. 2s Electrostatic potential energy profile of the SrRuO$_3$/BaTiO3/n-SrTiO$_3$ for two opposite polarization orientations.

For the SrRuO$_3$/PbTiO$_3$/n-SrTiO$_3$ system, we similarly estimate the screening potential at the PbTiO$_3$/n-SrTiO$_3$ interface to be about 0.57 eV. The electron affinity of PbTiO$_3$ is about 3.5 eV while the work function of n-SrTiO$_3$ is about 3.9 eV. Thus the barrier height between PbTiO$_3$ and n-SrTiO$_3$ is about 0.4 eV, which is smaller than the estimated screening potential 0.57 eV. Therefore, the conduction band minimum bends below the Fermi level for polarization pointing into the n-SrTiO$_3$ electrode and the interface layers of the PbTiO$_3$ gate metallize.

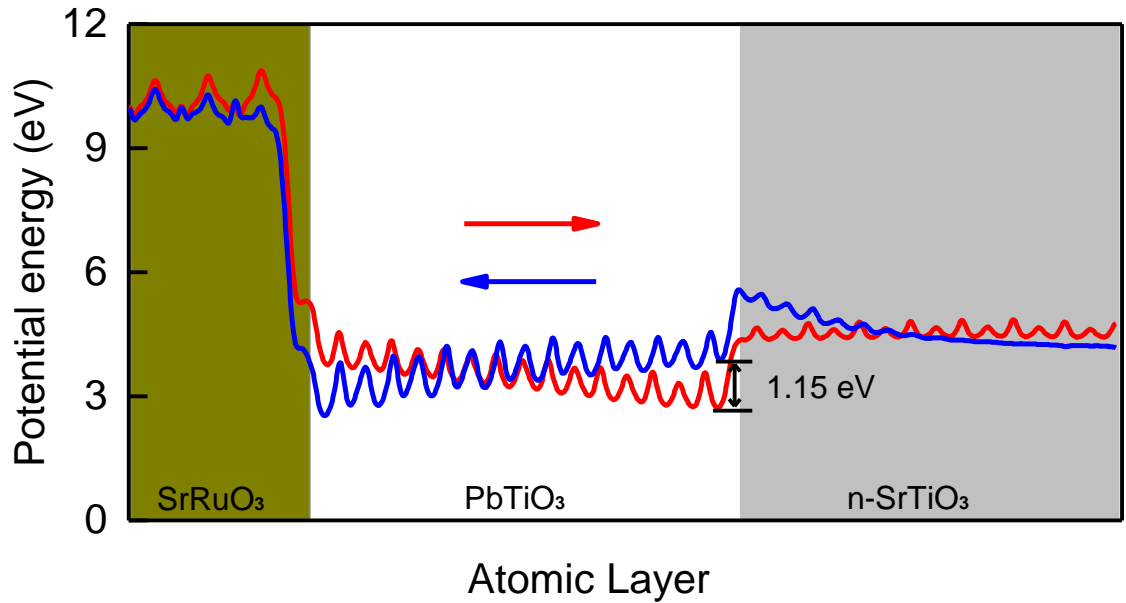

Fig. 3s Electrostatic potential energy profile of the SrRuO$_3$/PbTiO$_3$/n-SrTiO$_3$ for two opposite polarization directions.

**B: Calculations for SrRuO$_3$/PbTiO$_3$/n-SrTiO$_3$**

Our calculations on the PbTiO$_3$/n-SrTiO$_3$ interface indicates that about two unit cells of PbTiO$_3$ are metallized when the polarization in PbTiO$_3$ points into n-SrTiO$_3$, as shown by Fig. 4s:

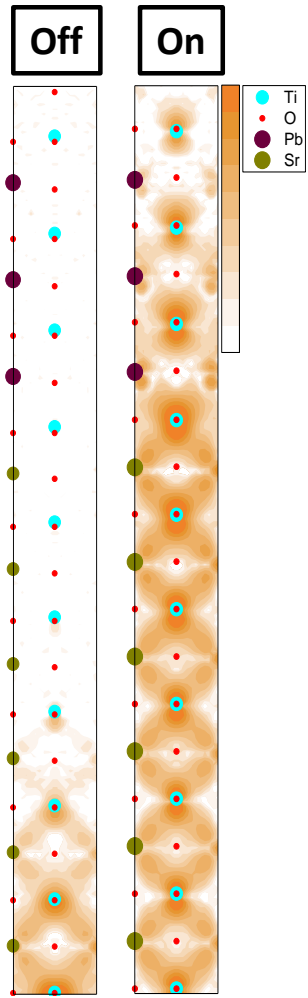

Fig. 4s. A 2D projection of the local density of electronic states, computed from first-principles calculations, integrated within $\pm k_B T$ eV of the Fermi level with T=300K near the PbTiO$_3$/n-SrTiO$_3$ interface.